\documentclass[twoside,twocolumn,9pt]{article}
\usepackage{extsizes}
\usepackage[super,sort&compress,comma]{natbib} 
\usepackage[version=3]{mhchem}
\usepackage[left=1.5cm, right=1.5cm, top=1.785cm, bottom=2.0cm]{geometry}
\usepackage{balance}
\usepackage{mathptmx}
\usepackage{sectsty}
\usepackage{graphicx} 
\usepackage{lastpage}
\usepackage[format=plain,justification=justified,singlelinecheck=false,font={stretch=1.125,small,sf},labelfont=bf,labelsep=space]{caption}
\usepackage{float}
\usepackage{fancyhdr}
\usepackage{fnpos}
\usepackage[english]{babel}
\addto{\captionsenglish}{%
}
\usepackage{array}
\usepackage{droidsans}
\usepackage{charter}
\usepackage[T1]{fontenc}
\usepackage{xcolor}
\usepackage{setspace}
\usepackage[compact]{titlesec}
 \usepackage{hyperref}
\usepackage{lmodern}
\usepackage{microtype}
\usepackage{hyperref}
\usepackage{graphicx}
\usepackage{amssymb}

\usepackage{epstopdf}

\definecolor{cream}{RGB}{222,217,201}

\begin{document}

\pagestyle{fancy}
\thispagestyle{plain}
\fancypagestyle{plain}{
\renewcommand{\headrulewidth}{0pt}
}

\makeFNbottom
\makeatletter
\renewcommand\LARGE{\@setfontsize\LARGE{15pt}{17}}
\renewcommand\Large{\@setfontsize\Large{12pt}{14}}
\renewcommand\large{\@setfontsize\large{10pt}{12}}
\renewcommand\footnotesize{\@setfontsize\footnotesize{7pt}{10}}
\makeatother

\renewcommand{\thefootnote}{\fnsymbol{footnote}}
\renewcommand\footnoterule{\vspace*{1pt}%
\color{cream}\hrule width 3.5in height 0.4pt \color{black}\vspace*{5pt}} 
\setcounter{secnumdepth}{5}

\makeatletter 
\renewcommand\@biblabel[1]{#1}            
\renewcommand\@makefntext[1]%
{\noindent\makebox[0pt][r]{\@thefnmark\,}#1}
\makeatother 
\renewcommand{\figurename}{\small{Fig.}~}
\sectionfont{\sffamily\Large}
\subsectionfont{\normalsize}
\subsubsectionfont{\bf}
\setstretch{1.125} 
\setlength{\skip\footins}{0.8cm}
\setlength{\footnotesep}{0.25cm}
\setlength{\jot}{10pt}
\titlespacing*{\section}{0pt}{4pt}{4pt}
\titlespacing*{\subsection}{0pt}{15pt}{1pt}

\fancyfoot{}
\fancyhead{}
\renewcommand{\headrulewidth}{0pt} 
\renewcommand{\footrulewidth}{0pt}
\setlength{\arrayrulewidth}{1pt}
\setlength{\columnsep}{6.5mm}
\setlength\bibsep{1pt}

\makeatletter 
\newlength{\figrulesep} 
\setlength{\figrulesep}{0.5\textfloatsep} 

\newcommand{\topfigrule}{\vspace*{-1pt}%
\noindent{\color{cream}\rule[-\figrulesep]{\columnwidth}{1.5pt}} }

\newcommand{\botfigrule}{\vspace*{-2pt}%
\noindent{\color{cream}\rule[\figrulesep]{\columnwidth}{1.5pt}} }

\newcommand{\dblfigrule}{\vspace*{-1pt}%
\noindent{\color{cream}\rule[-\figrulesep]{\textwidth}{1.5pt}} }

\makeatother

\twocolumn[
  \begin{@twocolumnfalse}

\vspace{1em}
\sffamily

{\centering
\noindent\LARGE{\textbf{Charge-partition pathways in strong-field photoionization of carbonyl sulfide monomers and dimers}} \\
\vspace{0.3cm}

\noindent\large{Chao He, Xinyue Zhang, Cangtao Yin, Markus Meuwly and Stefan Willitsch$^{\ast}$} \\
\vspace{0.3cm}
}

\noindent\normalsize{Strong-field photoionization of molecules and molecular clusters gives rise to a rich variety of fragmentation pathways governed by charge localization and redistribution on ultrafast timescales. Here, we report a velocity-map imaging study of the strong-field photoionization and fragmentation of carbonyl sulfide (OCS) monomers and dimers driven by 150 femtosecond (fs) laser pulses at 775~nm. The images of the total kinetic-energy and angular distributions of  the OCS$^{2+}$, S$^+$, and CO$^+$ fragments were interpreted with the help of electronic-structure calculations of the potential energy surfaces for OCS$^+$ and OCS$^{2+}$. We identify distinct dissociation pathways of singly and doubly ionized OCS, including two-body breakup channels of OCS$^+$ into $\mathrm{S}^+ +  \mathrm{CO}$ and $\mathrm{CO}^+ + \mathrm{S}$, dissociation of OCS$^{2+}$ into $\mathrm{S}^+ + \mathrm{CO}$$^+$ as well as higher-order three-body fragmentation. In addition, the images of the OCS$^{2+}$ channel exhibit near-zero-momentum components, low-energy isotropic features, and highly anisotropic contributions at high kinetic energies that cannot be explained by monomer ionization alone. Analysis of the KER distributions and angular anisotropies indicates that these features originate from the breakup of multiply charged OCS dimers ((OCS)$_2^{2+}$, (OCS)$_2^{3+}$, and (OCS)$_2^{4+}$) through charge-separation channels. Our results illustrate how dynamic signatures of strong-field fragmentation evolve from intramolecular dissociation in isolated molecules to intermolecular charge separation in weakly bound clusters providing a unified picture of charge-driven dissociation dynamics beyond the single-molecule limit.} \\

 \end{@twocolumnfalse} \vspace{0.6cm}
]

\renewcommand*\rmdefault{bch}\normalfont\upshape
\rmfamily
\section*{}
\vspace{-1cm}

\footnotetext{\textit{Department of Chemistry, University of Basel, Klingelbergstrasse 80, Basel 4056, Switzerland; E-mail: stefan.willitsch@unibas.ch}}



\section{Introduction}
Intense femtosecond laser fields can drive molecules far from equilibrium enabling access to highly excited electronic states and inducing multiphoton ionization on ultrafast timescales.\cite{kunitski2015observation, krausz2009attosecond, hishikawa2007visualizing, stapelfeldt2003colloquium} When several electrons are removed within a single laser pulse, the resulting Coulomb repulsion between positively charged fragments can dominate the ensuing nuclear motions leading to rapid fragmentation commonly referred to as Coulomb explosion.\cite{bocharova2011charge, stapelfeldt2003colloquium, ullrich2003recoil, posthumus2004dynamics} The kinetic energy and angular distributions of the resulting fragments carry detailed information about charge localization, the timing of ionization events, and the molecular geometry at the instant of fragmentation.\cite{kunitski2015observation, hishikawa2007visualizing, bocharova2011charge, stapelfeldt2003colloquium, ullrich2003recoil, seideman2002time} Velocity-map imaging (VMI)\cite{eppink1997velocity} represents a powerful experimental technique for the investigation of these processes by enabling measurements of fragment kinetic energies and angular distributions.\cite{eppink1997velocity, schouder2022laser, lin2003application} In particular, this method can elucidate distinct momentum-space signatures associated with different charge states and charge-separation pathways. 

To date, strong-field multiple ionization has been extensively characterized in isolated molecules,\cite{crane2023molecular} where charge is confined within a single molecular framework.\cite{posthumus2004dynamics, seideman2002time} Extending this framework to weakly bound molecular clusters reveals additional layers of complexity arising from intermolecular couplings.\cite{schouder2022laser} These interactions enable charge redistribution across the constituent moieties, giving rise to intermolecular charge separation, delayed ionization pathways, and fragmentation mechanisms absent in isolated molecules.\cite{siedschlag2004small} 

Carbonyl sulfide (OCS) represents an attractive model system for investigating these questions. As a linear triatomic molecule with a well-characterized electronic structure,\cite{brites2008ocs2+} OCS has been extensively studied under strong-field conditions,\cite{zhao2019strong} and its singly and doubly charged ions exhibit distinct fragmentation pathways.\cite{zhao2021tracking} At the same time, OCS readily forms van-der-Waals dimers and small clusters in supersonic expansions\cite{li2022ultrafast} offering a platform for exploring how weak intermolecular interactions modify strong-field ionization and fragmentation dynamics. The relatively simple molecular structure and dissociation channels of OCS further facilitate the interpretation of momentum-resolved measurements.

Masuoka and coworkers established a comprehensive characterization of OCS fragmentation following multiple photoionization in the 20–100 eV photon energy range by employing photoelectron–photoion–photoion coincidence (PEPIPICO) techniques in combination with synchrotron radiation.\cite{masuoka1991dissociative, masuoka1992kinetic, masuoka1993single, masuoka1991anisotropic, masuoka1993dissociation} These studies provided detailed insights into the dissociation dynamics of singly, doubly, and triply charged OCS species and yielded dissociative-photoionization cross sections and appearance energies for the dominant fragmentation channels. For singly charged \ce{OCS+}, the principal dissociation pathways were identified as \ce{CO + S+}, \ce{CO+ + S}, \ce{CS+ + O}, and \ce{CS + O+}.
While \ce{OCS^2+} may remain metastable near its appearance threshold of approximately 31 eV, at higher photon energies it predominantly dissociates in a combination of two-body channels, \ce{CO+ + S+} and \ce{O+ + CS+}, and three-body fragmentation pathways proceeding either sequentially or concertedly to form \ce{C+ + S+ + O}, \ce{O+ + S+ + C}, and \ce{C+ + O+ + S}. 
At still higher photon energies, triple photoionization of OCS leads to rapid dissociation of \ce{OCS^3+}, with the dominant channels being a concerted three-body breakup into \ce{C+ + O+ + S+} and two-body charge separation into \ce{CO+ + S^2+}.

Subsequent ultrafast strong-field-ionization studies have provided deeper insights into the dynamics of these fragmentation pathways. Zhao et al. employed time-resolved Coulomb-explosion imaging to probe the strong-field-induced dissociation of OCS,\cite{zhao2019strong} resolving both two-body and three-body fragmentation channels. 
Complementary work by Endo \textit{et al.} focused on the fragmentation of doubly ionized OCS produced by strong-field double ionization using two-color femtosecond laser fields and coincidence momentum imaging.\cite{endo2022post} In this case, \ce{CO+ + S+} was identified as the dominant two-body channel.

Ding and co-workers presented a refined and comprehensive four-body coincidence Coulomb-explosion imaging study to uncover isomers of carbonyl sulfide dimers, \ce{(OCS)2}.\cite{zhao2021tracking, yu2021determining, li2022ultrafast} In their work, the dimers were multiply ionized by an intense 
circularly polarized laser pulse, triggering the four-body fragmentation pathway \ce{(OCS)2^{4+} \rightarrow CO+ + S+ + CO+ + S+}. The three-dimensional momenta of all four ionic fragments were measured in coincidence using a cold target recoil ion momentum spectrometer (COLTRIMS). By performing kinetic-energy correlation analyses in conjunction with classical Coulomb explosion simulations, these authors successfully disentangled sequential fragmentation from concerted breakup processes. Four distinct isomers of the dimer were clearly identified under their experimental conditions.  

More recently, Lomas \textit{et al.} investigated the dissociative-electron-impact-ionization dynamics of OCS using multimass velocity-map and covariance-map imaging supported by quantum-chemical calculations of the corresponding potential energy surfaces (PESs) for OCS$^+$ and OCS$^{2+}$.\cite{lomas2024multimass} Their findings are consistent with the previous photoionization work and lend further support to the conclusions of those studies.

Here, we present a VMI study of strong-field ionization and fragmentation of OCS and OCS dimers using 150 fs laser pulses at 775 nm. By imaging the kinetic-energy and angular distributions of OCS$^{2+}$, S$^+$ and CO$^+$ species, supported by calculated PESs for OCS$^+$ and OCS$^{2+}$, we identify multiple, distinct dissociation pathways that arise from different charge-partition mechanisms in monomeric and clustered environments. Through VMI of \ce{CO+} and \ce{S+} fragments at varying laser intensities, we disentangle distinct dissociation pathways of singly and doubly ionized OCS. Our results reveal two-body fragmentation channels of \ce{OCS+} leading to \ce{CO+ + S} and \ce{S+ + CO}, as well as Coulomb-repulsion-driven dissociation of OCS$^{2+}$ into \ce{CO+ + S+} accompanied by higher-order three-body fragmentation processes. 

The images of OCS$^{2+}$ reveal coexisting near-zero-momentum components, low-energy isotropic features, and highly anisotropic contributions extending to electron-volt energies, which cannot be explained by the ionization of isolated molecules alone. Our analysis indicates that these features arise from the fragmentation of multiply charged OCS dimers via distinct charge-separation pathways, namely $(\mathrm{OCS})_2^{2+} \rightarrow \mathrm{OCS}^{2+} + \mathrm{OCS}$, $(\mathrm{OCS})_2^{3+} \rightarrow \mathrm{OCS}^{2+} + \mathrm{OCS}^{+}$, and $(\mathrm{OCS})_2^{4+} \rightarrow \mathrm{OCS}^{2+} + \mathrm{OCS}^{2+}$. The simultaneous presence of high- and intermediate-kinetic-energy features in the images reflects competing ionization channels and illustrates that the present strong-field ionization of weakly bound dimers is controlled by intermolecular charge-partition dynamics. The current study thus addresses the strong-field ionization of OCS monomers and dimers within a unified experimental framework enabling a direct comparison that elucidates how charge confinement in isolated molecules gives way to intermolecular charge redistribution in weakly bound aggregates.

\begin{figure*}
 \centering
 \includegraphics[height=5cm]{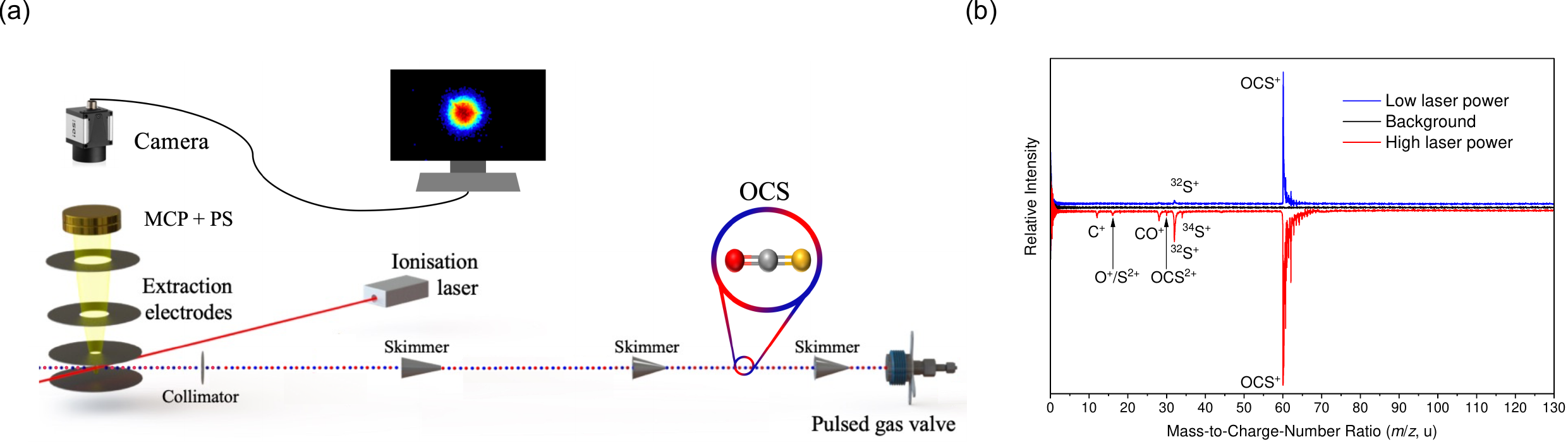}
 \caption{(a) Schematic illustration of the molecular-beam apparatus used in the present experiments. (b) Time-of-flight mass spectra of the ionic products generated by strong-field ionization of 1000 ppm carbonyl sulfide (OCS) seeded in Helium (He) at high (red inverted trace) and low (blue trace) laser intensity. The black trace corresponds to the background signal. See text for details.}
 \label{fgr:Figure 1}
\end{figure*}
\section{Methods}

\subsection{Experiment}
The experiments were carried out using a molecular-beam setup, the details of which have been described previously.\cite{ploenes2024collisional,mishra2025isomer} A gas mixture of 1000 ppm carbonyl sulfide (OCS, Aldrich, $\ge 97.5$\%) seeded in helium (PanGas, 99.996\%) at a stagnation pressure of 9 bars was supersonically expanded via a pulsed Even-Lavie valve (150 $\mu\mathrm{m}$ orifice) \cite{even2000cooling} into a source vacuum chamber and subsequently collimated by skimmers (Figure~\ref{fgr:Figure 1}(a)). In the photoionization chamber, the pulsed molecular beam traveled perpendicular to the axis of a time-of-flight (TOF) mass spectrometer (MS) and intersected with the femtosecond (fs)-laser beam (Clark-MXR, Inc. 775 nm, 150 fs) beam at 45° angle between the repeller and the extraction plates of the VMI ion optics. The fs-laser beam was focused into the VMI region using a 70 mm focal-length lens.

Using a set of ion optics in VMI configuration \cite{eppink1997velocity,xu2024trapping}, the product ions generated by multiphoton ionization were projected onto a 75 mm dual multichannel plate (MCP, Photek) detector paired with a P46 phosphor screen (PS) to image their kinetic-energy and angular distributions. For TOF-MS measurements, the output from the MCP was amplified using a home-built x10 preamplifier prior to being digitized by an oscilloscope (Teledyne LeCroy WaveRunner 8054). To achieve mass selection and temporal gating in the images, a fast high-voltage pulse (Photek, GM-MCP3 Gating Module, pulse amplitude of 950 V and pulse duration of 40 ns) was applied to one of the MCPs, performing what is commonly referred to as DC slicing of the ion packet in VMI measurements.\cite{lin2003application} The images on the phosphor screen were recorded with a fast CMOS camera (IDS UI-3040CP-M-GL Rev.2) and subsequently sent to a computer after every laser shot for ion-event detection and further data analysis.\cite{mishra2025isomer} Each raw image was typically obtained by integrating a minimum of 9 million laser shots. Corresponding background images were collected under identical conditions with the molecular beam switched off. A repetition rate of 50 Hz was used for all measurements. The timings of the pulsed valve, the pulsed laser, and the pulsed gating of the imaging detector were synchronized by a PulseBlasterESR-PRO delay generator (Spin-Core Technologies).

\subsection{Theoretical calculations}

Single-point electronic-structure calculations were carried out using a multireference approach implemented in the MOLPRO program package.\cite{molpro2020} For neutral OCS, multi-configuration self-consistent-field (MCSCF)\cite{werner1985second,knowles1985efficient,kreplin2019second,kreplin2020mcscf} calculations with an active space of 12 electrons in 13 orbitals (12e,13o) were followed by multi-reference configuration-interaction (MRCI)\cite{werner1988efficient,knowles1988efficient,knowles1992internally} calculations with a reduced active space of (12e,10o). For OCS$^+$, MCSCF(11e,13o) calculations were followed by MRCI(11e,10o), and for OCS$^{2+}$, MCSCF(10e,13o) calculations were followed by MRCI(10e,10o). In all cases, the Davidson relaxed reference correction\cite{langhoff1974configuration} was applied. The aug-cc-pVTZ (aVTZ) basis set was used throughout this work.

Franck–Condon (FC) factors were computed along the normal coordinate of the C-S stretching mode. Potential-energy curves for each electronic state were obtained by varying this coordinate and calculating the corresponding electronic energies, which were interpolated using cubic splines to yield smooth potentials. Vibrational wavefunctions were determined by numerically solving the one-dimensional Schrödinger equation using a finite-difference approach.\cite{Thijssen2007computational} FC factors were then calculated as squared overlaps between vibrational wavefunctions of the initial and final states. 

\section{Results and discussion}
\subsection{Time-of-flight mass spectra}

A representative TOF mass spectrum recorded under low ($(1.9 \pm 0.2)\times 10^{13}\ \mathrm{W\,cm^{-2}}$) and high ($(3.2 \pm 0.2)\times 10^{13}\ \mathrm{W\,cm^{-2}}$) laser intensities for multiphoton ionization is shown in Figure~\ref{fgr:Figure 1}(b). Background TOF mass spectra were recorded under identical conditions with the OCS molecular beam turned off. No background signals were detected at any laser power, confirming that the observed photoionization signals arise solely from OCS-derived products. A representative background spectrum acquired at the high laser power is shown in Figure~\ref{fgr:Figure 1}(b). At low laser power, only two ion signals were detected: ${m/z}$ = 60~u (OCS$^+$) and ${m/z}$ = 32~u ($^{32}$S$^+$). Among these, OCS$^+$ appears as the dominant product, while S$^+$ is present only as a minor contribution. 

As the laser power was increased, the intensities of ${m/z}$ = 60~u (OCS$^+$) and 32~u ($^{32}$S$^+$) peaks increased correspondingly, indicating enhanced ion formation. Under these conditions, additional product ions were observed in the mass channels ${m/z}$ = 30~u (OCS$^{2+}$), 28~u (CO$^+$), 16~u (O$^+$/S$^{2+}$), and 12~u (C$^+$). At the present dilution, the OCS/He molecular beam contained not only monomers, but also a small, yet non-negligible fraction of dimers and larger clusters. This is evidenced at even higher laser power ($(6.0 \pm 0.2)\times 10^{13}\ \mathrm{W\,cm^{-2}}$) at which a distinct (OCS)$_2^+$ signal (${m/z}$ = 120~u) was observed (Figure S1 of the ESI). At all laser powers used, the OCS$^+$ was found to be the dominant product ion. In the relevant images (not shown), it appeared as a compact central spot which indicates photoionization from the parent neutral in the beam and reveals no further dynamics. The following analysis thus focuses on the results of the imaging of OCS$^{2+}$, S$^+$, and CO$^+$ performed at low ($(1.9 \pm 0.2)\times 10^{13}\ \mathrm{W\,cm^{-2}}$) and high ($(3.2 \pm 0.2)\times 10^{13}\ \mathrm{W\,cm^{-2}}$) laser intensities.

\begin{figure*}
 \centering
 \includegraphics[height=7cm]{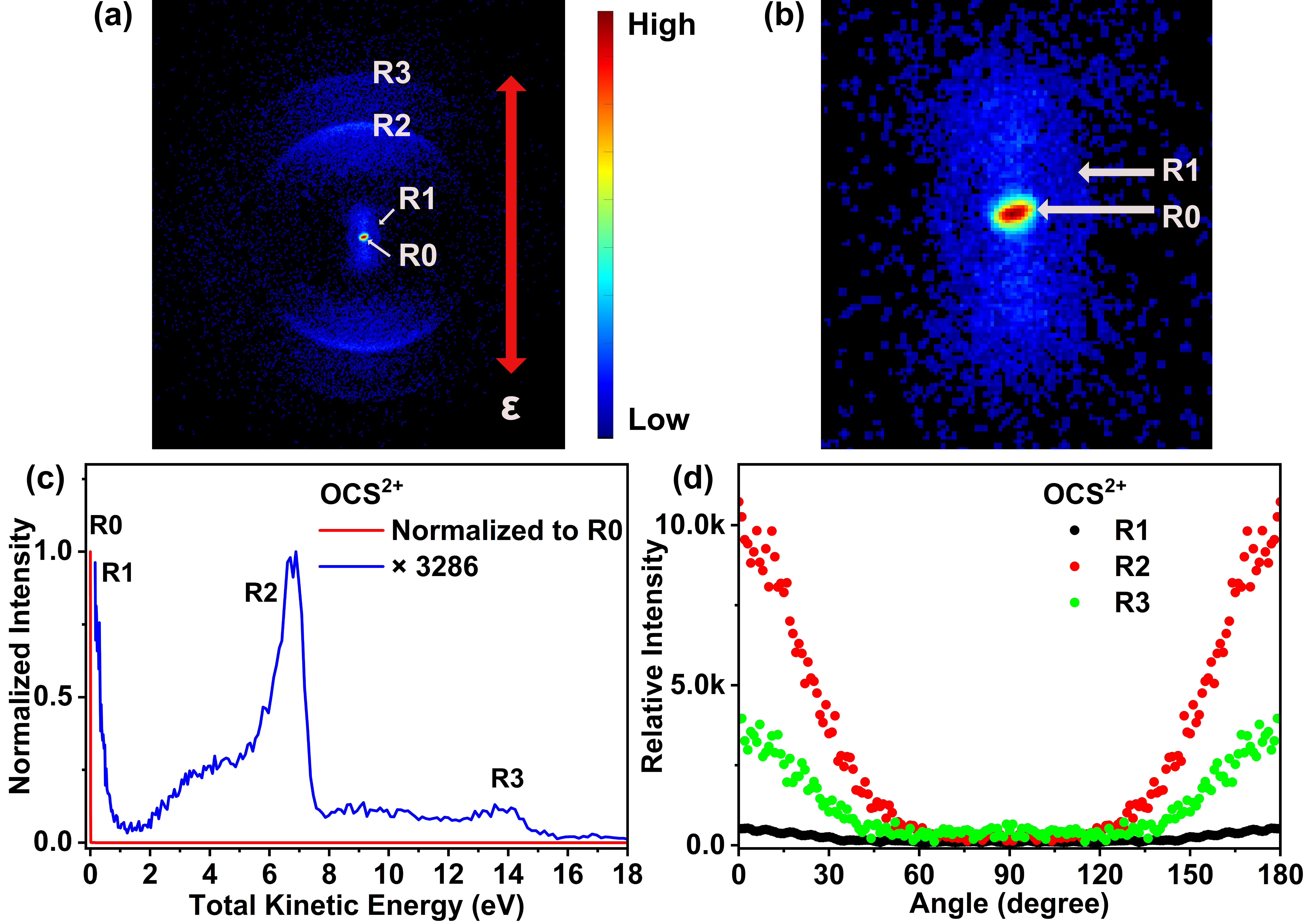}
 \caption{(a) Velocity-mapped ion image obtained in the OCS$^{2+}$ channel. A logarithmic color 
scale is used to enhance the visibility of the weaker features. (b) Magnified central part of the image. (c) Total kinetic-energy distributions of OCS$^{2+}$ extracted from the image. (d) Angular distribution of product channels R1, R2, R3 of OCS$^{2+}$. The orientation of the polarization vector $\epsilon$ of the laser radiation is shown by the double-headed red arrow.}
 \label{fgr:Figure 2}
\end{figure*}

\begin{table}[h]
\small
  \caption{\ Minimum energies and equilibrium bond lengths $r$ of electronic states and dissociation asymptotes of OCS$^+$ and OCS$^{2+}$, relative to OCS(${\mathrm{X}}^1\Sigma^+$)}
  \label{tbl:minima}
  \begin{tabular*}{0.48\textwidth}{@{\extracolsep{\fill}}lllll}
    \hline
    State & $E$ (eV) & Expt.\cite{eland1973predissociation,wang1988high,brites2008ocs2+} & $r_{\mathrm{CO}}$ (Å) & $r_{\mathrm{CS}}$ (Å) \\
    \hline
    OCS$^+$(${\mathrm{X}}^2\Pi$) & 11.0 & 11.2 & 1.13 & 1.63 \\
    OCS$^+$(${\mathrm{A}}^2\Pi$) & 14.9 & 15.1 & 1.19 & 1.62 \\
    OCS$^+$(${\mathrm{B}}^2\Sigma^+$) & 15.6 & 16.0 & 1.16 & 1.59 \\
    CO(${\mathrm{X}}^1\Sigma^+$)+S$^+$($^4$S$_u$) & 13.1 & 13.5  \\
    CO(${\mathrm{X}}^1\Sigma^+$)+S$^+$($^2$D$_u$) & 14.9 & 15.3 \\
    CO(${\mathrm{X}}^1\Sigma^+$)+S$^+$($^2$P$_u$) & 16.2 & 16.6 \\
    OCS$^{2+}$(${\mathrm{X}}^3\Sigma^-$) & 29.6 & 30.1 & 1.13 & 1.70 \\
    OCS$^{2+}$(${\mathrm{a}}^1\Delta$) & 30.8 & 31.1 & 1.14 & 1.68 \\
    OCS$^{2+}$(${\mathrm{b}}^1\Sigma^+$) & 31.7 & 31.8 & 1.16 & 1.65 \\
    \hline
  \end{tabular*}
\end{table}

\subsection{OCS\texorpdfstring{$^{2+}$}{2+}}
The OCS$^{2+}$ product image recorded at high laser power is presented in Figure~\ref{fgr:Figure 2}, where four distinct features corresponding to different dissociation pathways—labelled R0, R1, R2, and R3—are identified (Figure~\ref{fgr:Figure 2}(a,b)). The image radius was converted into the kinetic energy of the detected OCS$^{2+}$ fragment and subsequently into the corresponding total kinetic-energy release (KER) for the associated two-body dissociation channel using momentum and energy conservation, accounting for the mass of the undetected co-fragment. The resulting KER distribution is shown in Figure~\ref{fgr:Figure 2}(c). Because the intensities of R1–R3 are much weaker than that of R0, normalization to R0 (red trace in Figure~\ref{fgr:Figure 2}(c)) renders the R1–R3 features barely discernible. To illustrate their spatial distributions, the intensities of R1–R3 (blue trace in Figure~\ref{fgr:Figure 2}(c)) were therefore scaled by normalizing the R2 feature to 1. Consequently, the image in Figure~\ref{fgr:Figure 2}(a) is displayed on a logarithmic scale to enhance the visibility of the R1–R3 features. R0 appears as an intense, central spot corresponding to a near-zero KER. In contrast, R1, R2, and R3 form well-defined concentric rings with characteristic kinetic energies of approximately 0.06~eV, 3.5~eV, and 6.98~eV, respectively (Figure~\ref{fgr:Figure 2}(c)). These features exhibit similar angular distributions, with intensity maxima aligned along the laser-polarization axis (Figure~\ref{fgr:Figure 2}(d)).

\begin{figure}[h]
\centering
  \includegraphics[height=13cm]{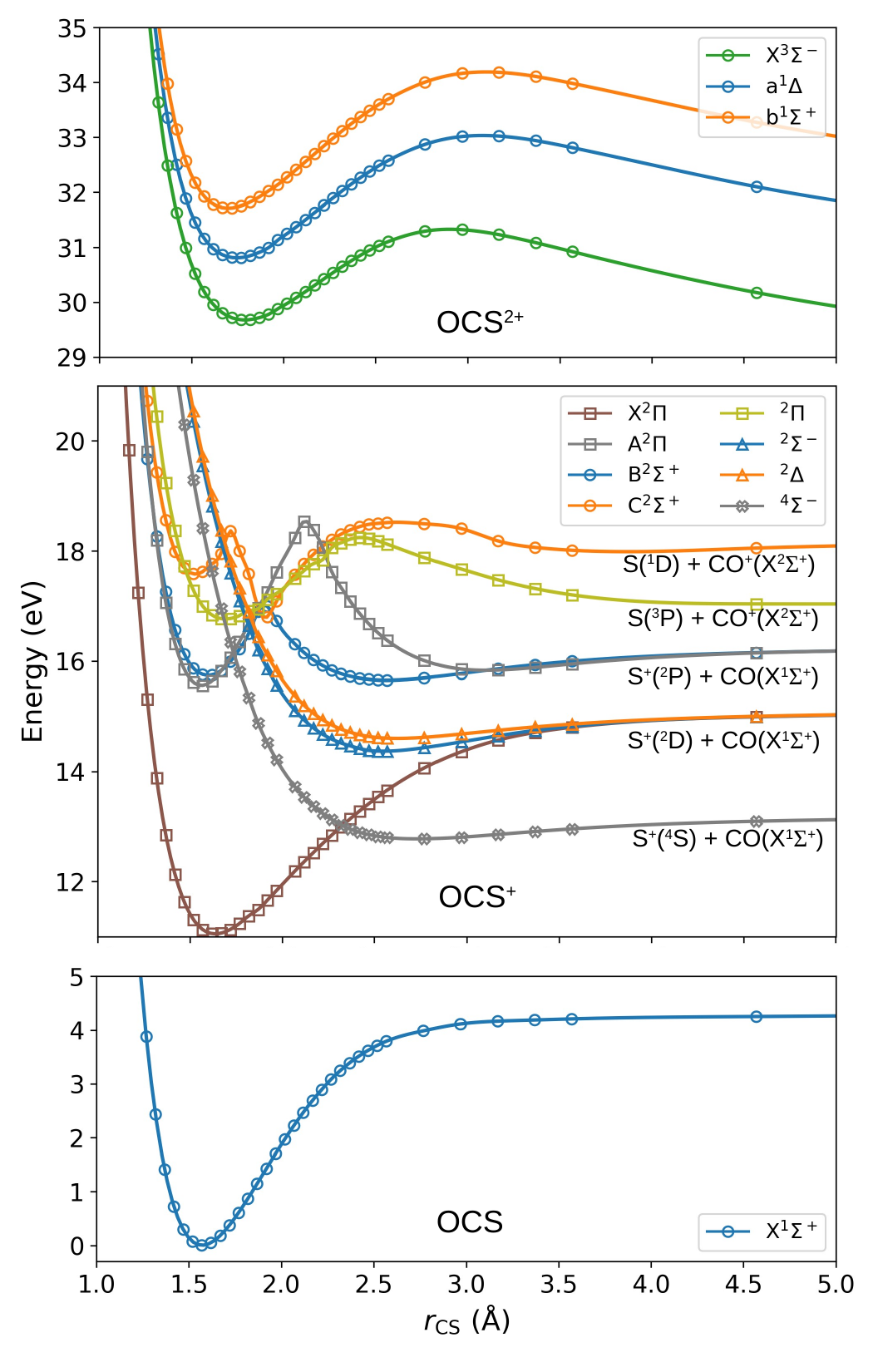}
  \caption{Potential energy curves for the dissociation of OCS, OCS$^+$, and OCS$^{2+}$ computed along the C–S bond coordinate at the MRCI/aVTZ level. The molecule was constrained at a linear geometry with the C–O bond length fixed at 1.158 \AA, corresponding to the equilibrium value of OCS.}
  \label{fgr:Figure 3}
\end{figure}

\subsubsection{R0 channel: population of the lowest bound states of OCS\texorpdfstring{$^{2+}$}{2+}}

The three lowest electronic states of OCS$^{2+}$ ($X^3 \Sigma^-$, $a^1 \Delta$, and $b^1 \Sigma^+$) feature deeply bound wells (see PES in Figure~\ref{fgr:Figure 3} and Table~\ref{tbl:minima}). They are stable against prompt fragmentation\cite{endo2022post} and can be accessed within the Franck-Condon region for excitation from the vibronic ground state of neutral OCS, see Table~S1 and Figure S2 of the ESI. The formation and stability of these states of OCS$^{2+}$ have been investigated previously using threefold and fourfold electron–ion coincidence spectroscopy supported by quantum-chemical calculations on their different isomeric structures.\cite{jarraya2023doubly} 
Photoionization spectra recorded in coincidence with the parent dication showed three characteristic bands indicating population of the  $X^3 \Sigma^-$, $a^1 \Delta$, and $b^1 \Sigma^+$ states. However, the parent OCS$^{2+}$ signal was found to disappear at excitation energies above approximately 34~eV indicating access to higher-lying dissociative states of OCS$^{2+}$. These previous findings suggest that the R0 feature observed in the image of Figure~\ref{fgr:Figure 2} arises from population of the lowest three electronic states of OCS$^{2+}$. The corresponding ionization potentials calculated here - 29.6 eV, 30.8 eV, and 31.7 eV (Table~\ref{tbl:minima}) - agree with the previously computed values of 30.1 eV, 31.1 eV, and 31.8 eV,\cite{jarraya2023doubly,brites2008ocs2+}. Population of these states using 775 nm radiation proceeds via multiphoton absorption processes involving $\gtrsim$19 photons. 

\subsubsection{R1–R3 channels: fragmentation of OCS clusters.~~} 

Although OCS monomers dominate the molecular beam under the current experimental conditions, a non-negligible population of dimers and larger clusters was also present. The KER distributions observed for features R1, R2, and R3 in Figure~\ref{fgr:Figure 2} suggest that these signals arise from fragmentation of higher-mass species, most plausibly OCS dimers.\cite{schouder2022laser} The appearance of these features in the OCS$^{2+}$ mass channel implies that they must originate from at least doubly ionized dimers, \ce{(OCS)2^2+}, and possibly from even higher charge states such as \ce{(OCS)2^3+} and \ce{(OCS)2^4+}.\cite{li2022ultrafast} However, the probability of forming such highly charged species decreases rapidly with increasing charge state. 

Assuming a breakup into charged moieties where each fragment can be modeled as a point charge, the Coulomb potential energy $U$ stored in a dimer at the moment of ionization is given by

\begin{equation}
U = \frac{q_1 q_2}{4\pi \varepsilon_0 r},
\end{equation}

where $r$ is the separation between the charges $q_1$ and $q_2$ on the fragments, and \ce{\varepsilon0} is the vacuum permittivity.\cite{pickering2018femtosecond} Further assuming that this potential energy is fully converted into kinetic energy during such a Coulomb-driven dissociation, the associated kinetic-energy release (KER) can be estimated for different charge configurations.\cite{song2022dissociative} 

Assuming that channels R2 and R3 in Figure~\ref{fgr:Figure 2} arise from the fragmentation of OCS dimers, the relevant total KERs are determined to be 7.0 eV and 13.96 eV (Figure~\ref{fgr:Figure 2} (c)). Note that channel R3 exhibits nearly double the KER of R2. This observation suggests the assignment of R2 to the fragmentation of a triply charged moiety, i.e., \ce{(OCS)2^3+ -> OCS^2+ + OCS+}, and of R3 to \ce{(OCS)2^4+ -> OCS^2+ + OCS^2+}. From the measured KERs, thebond lengths of laser-produced \ce{(OCS)2^3+} and \ce{(OCS)2^4+} ions are estimated to be 4.11 Å and 4.13 Å, respectively. These values fall well within the range of $\approx 4-5$~Å reported for the bond lengths of different neutral OCS dimers.\cite{yu2021determining, brown2012computational} 

By contrast, channel R1 is characterized by a markedly lower KER, extending only to approximately 0.5~eV. In particular, this value lies well below the range expected for Coulomb-driven fragmentation, suggesting a fundamentally different dissociation pathway. We therefore attribute this channel to dissociation of doubly ionized OCS dimers via \ce{(OCS)2^2+ -> OCS^2+ + OCS}, in which a neutral OCS fragment is formed. Notably, the R1 feature consists of two distinct components, see Figure~\ref{fgr:Figure 2} (b): an isotropic ring centered at a kinetic energy of approximately 0.06~eV, and an anisotropic contribution whose intensity is enhanced along the laser polarization axis and extends to higher kinetic energies up to approximately 0.25 eV. Given the short duration of the ionizing laser pulses, double ionization is expected to occur on a timescale much shorter than nuclear motion, resulting in the preparation of highly vibrationally excited \ce{(OCS)2^2+} species. The subsequent dissociation dynamics are therefore likely governed by intramolecular vibrational energy redistribution (IVR).\cite{nesbitt1996vibrational} If dissociation occurs promptly, i.e., IVR is faster than the timescale of molecular rotation, residual alignment imposed by the laser polarization can be preserved, giving rise to the observed anisotropic component. In contrast, slow dissociation leads to loss of directional memory and produces an isotropic component in the images. The coexistence of isotropic and anisotropic features observed for the R1 channel thus suggests the coexistence of fast and slow dissociation mechanisms.

\begin{figure*}
 \centering
 \includegraphics[height=7cm]{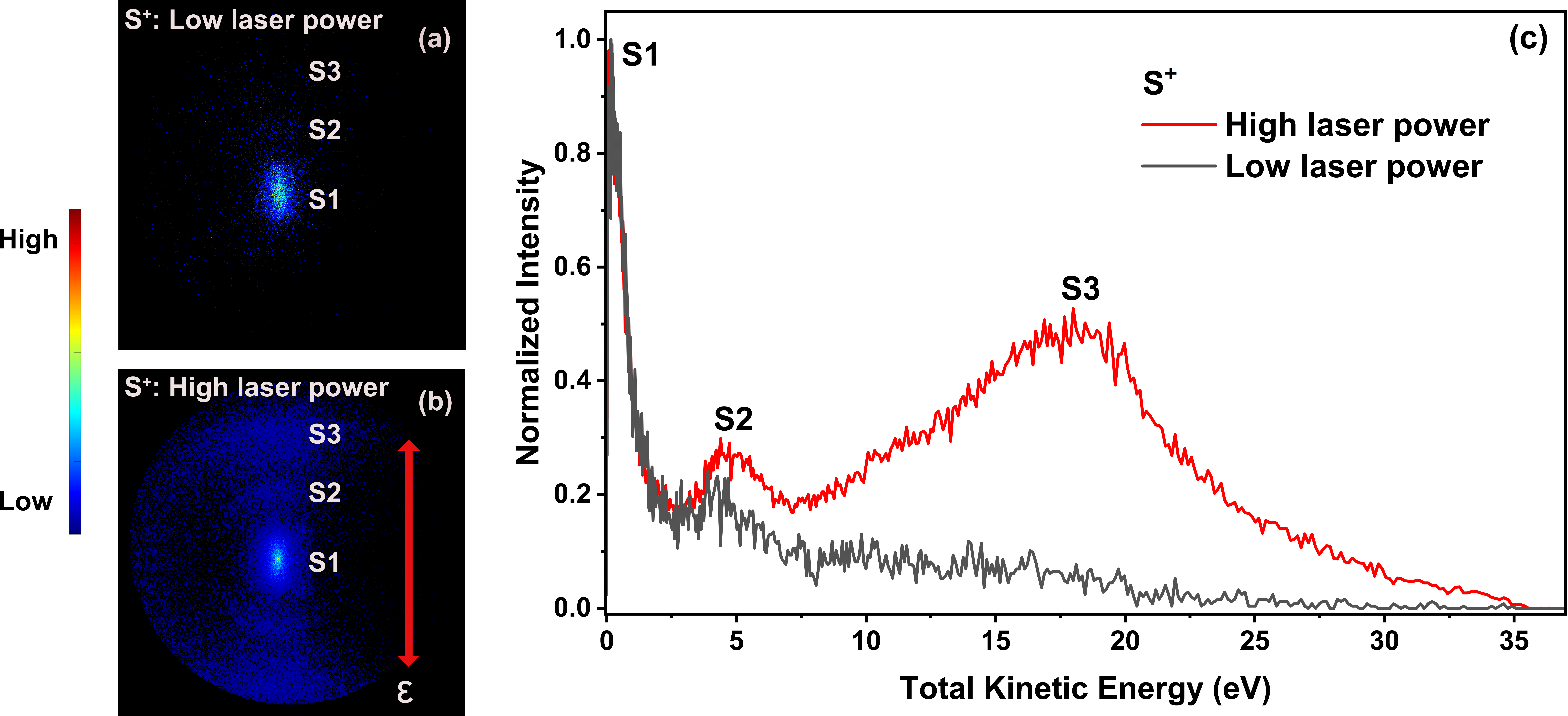}
 \caption{Images obtained at the mass of the S$^+$ product obtained at (a) low and (b) high laser power. The observed dissociation pathways are denoted as S1, S2, and S3. The orientation of the laser polarization vector ($\epsilon$) is indicated by the double-headed red arrow. (c) Corresponding total kinetic-energy distributions.}
 \label{fgr:Figure 4}
\end{figure*}

\subsection{S$^+$ channel}

\begin{figure}[h]
\centering
  \includegraphics[height=5.5cm]{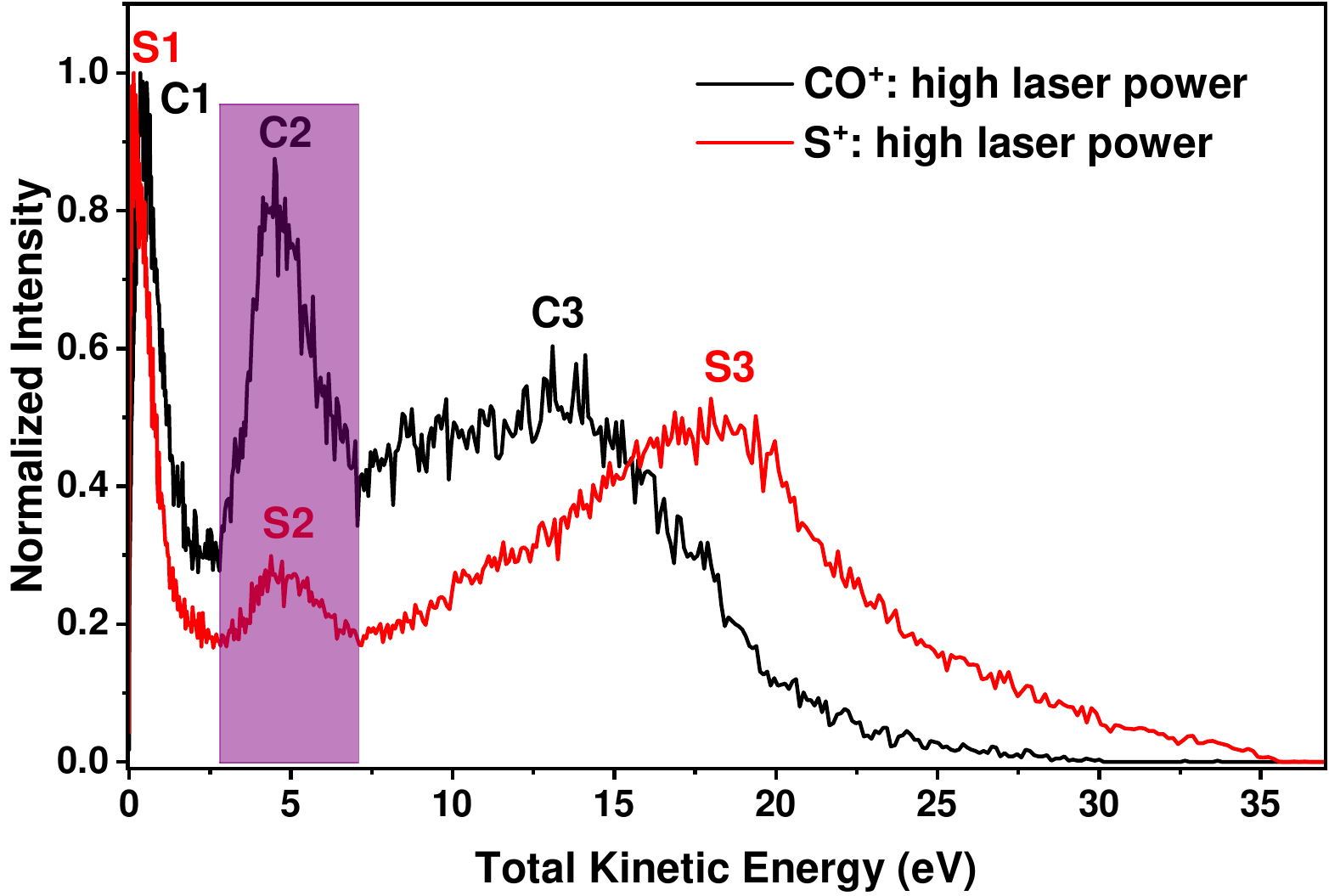}
  \caption{Total kinetic energy distributions that derived from S$^+$ and CO$^+$ imagings at high laser power.}
  \label{Figure 5}
\end{figure}

\begin{figure*}
 \centering
 \includegraphics[height=7cm]{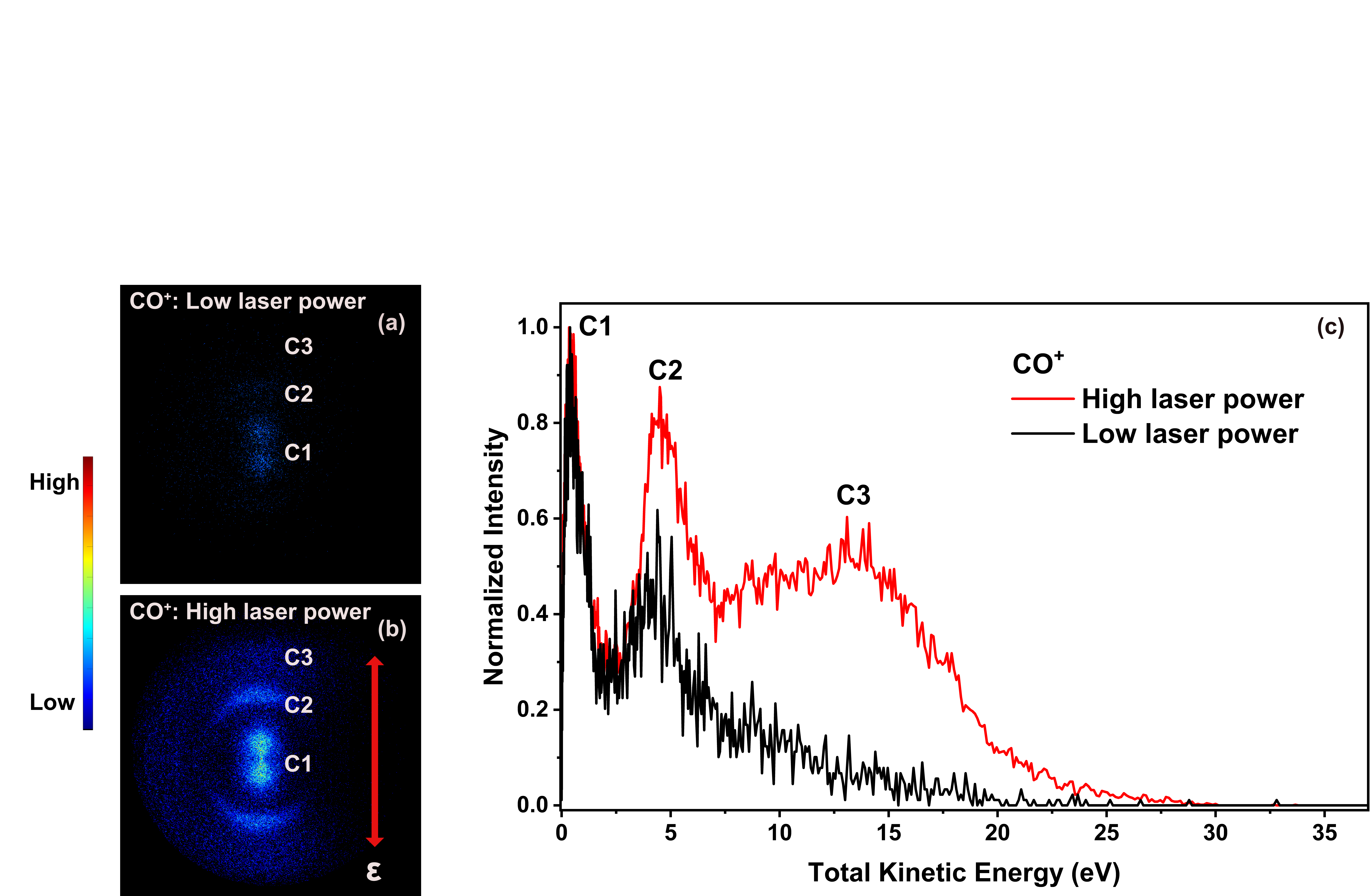}
 \caption{Images obtained at the mass of the CO$^+$ product obtained at (a) low and (b) high laser power. The observed dissociation pathways are denoted as C1, C2, and C3. The orientation of the laser polarization vector ($\epsilon$) is indicated by the double-headed red arrow. (c) Corresponding total kinetic-energy distributions.}
 \label{fgr:Figure 6}
\end{figure*}

The images recorded for the S$^+$ product mass at low and high laser powers are shown in Figure~\ref{fgr:Figure 4} (a) and (b), respectively. Three distinct features labeled as S1, S2, and S3 can be discerned. At low laser power, channel S1 is the dominant product, while S2 appears only as a minor contribution, and S3 is not observed. As the laser power is increased, S2 becomes more pronounced, and channel S3 emerges, indicating the activation of additional dissociation pathways under stronger laser fields. All three channels (S1, S2, and S3) exhibit anisotropic angular distributions (Figure S3 of the ESI), with the product flux preferentially aligned along the direction of the laser polarization. Given that the dominant fragmentation channels for OCS$^+$ are S$^+$ + CO and S + CO$^+$, while OCS$^{2+}$ predominantly dissociates into S$^+$ + CO$^+$, the kinetic energy distributions derived from the S$^+$ images (Figure~\ref{fgr:Figure 4} (c)) correspond to the total KER of the CO/CO$^+$ + S$^+$ fragment pairs. Likewise, the kinetic energy distributions derived from the CO$^+$ images (Figure~\ref{Figure 5} and Figure~\ref{fgr:Figure 6} (c)) correspond to the total KER of the CO$^+$ + S/S$^+$ dissociation channels.

For channel S1, the KER distribution spans from 0 eV to 2.5 eV, with a pronounced peak at approximately 0.14 eV. Upon ionization, OCS$^+$ may be populated in the $X^2 \Pi$, $A^2 \Pi$, and $B^2 \Sigma^+$ electronic states, each capable of dissociation to form $\mathrm{S}^+ + \mathrm{CO}$.\cite{wang2024high, wang2024vibrational, chang2005imaging} Potential-energy curves of the low-lying electronic states of OCS$^+$ are shown in Figure~\ref{fgr:Figure 3}. Photoelectron spectra of OCS \cite{wang1988high,holland1990photoelectron} showed distinct excitations to the $X^2 \Pi$, $A^2 \Pi$, $B^2 \Sigma^+$ and $C^2 \Sigma^+$ states. The photoelectron band corresponding to the $B$ state  which features almost diagonal Franck-Condon factors with the neutral vibrational ground state (Table~S1 in the ESI) is particularly pronounced and overlaps a progression of vibrational bands attributed to population of the $A$ state. 

A number of dissociation pathways must be considered for an interpretation of the S1 feature. In general, one can expect that the fragmentation dynamics depend strongly on the electronic state. The ground state of OCS$^+$($X^2 \Pi$) can dissociate directly to $\mathrm{S}^+(^2D) + \mathrm{CO}(X^1\Sigma^+)$ and may also couple to the $^4\Sigma^-$ state to yield $\mathrm{S}^+(^4S) + \mathrm{CO}(X^1\Sigma^+)$ (Figure~\ref{fgr:Figure 3}). However, Franck-Condon factors to vibrational levels of the X state energetically located above these dissociation asymptotes are expected to be small (Table~1 of the ESI). Thus, most of the population generated in the X state is expected to contribute to undissociated OCS$^+$ product. The $A^2\Pi$ state must undergo coupling to the $^2\Sigma^-$ and/or $^2\Delta$ states before producing $\mathrm{S}^+(^2D) + \mathrm{CO}(X^1\Sigma^+)$. Furthermore, population in the $A^2 \Pi$ or $B^2 \Sigma^+$ states may relax to the ground $X^2 \Pi$ state prior to dissociation. 

The present results can be compared with previous direct photodissociation studies of OCS$^+$. Liu and co-workers \cite{chang2005imaging} showed that predissociation of mode-selected $\mathrm{OCS}^{+}\!\left[(\nu_{1}\,\nu_{2}\,\nu_{3})\,B^{2}\Sigma^{+}\right]$ yields exclusively the $\mathrm{S}^+(^2D) + \mathrm{CO}(X^1\Sigma^+)$ channel. These observations were rationalized by competing nonadiabatic pathways involving internal conversion to the $X$ state and/or coupling to other electronic states via a number of energetically accessible conical intersections (see Figure~\ref{fgr:Figure 3}) prior to dissociation, with the relative importance of each pathway determined by the initially excited vibrational mode. 

Recently, systematic investigations of the photodissociation of OCS$^+$ prepared in selected vibrational levels of the $A^{2}\Pi$ state by Wang and co-workers\cite{wang2024high, wang2024vibrational} have shown that dissociation of OCS$^+$ proceeds via both the $\mathrm{S}^+(^4S) + \mathrm{CO}(X^1\Sigma^+)$ and $\mathrm{S}^+(^2D) + \mathrm{CO}(X^1\Sigma^+)$ asymptotes which was also rationalized in terms of internal conversion to the $X$ state and other non-adiabatic couplings. Thus, it appears plausible that the S1 feature in Figure~\ref{fgr:Figure 4} predominantly arises from dissociation of the $B$ state of OCS$^+$ and vibrational components of the $A$ state populated according to Franck-Condon factors (Table~S1 of the ESI) which predissociate to the two lowest asymptotes through non-adiabatic couplings.

The S1 channel exhibits a pronounced anisotropic angular distribution (Figure S3) along the direction of the laser polarization. This behavior is consistent with the established orientation dependence of strong-field ionization in OCS, where the ionization probability depends sensitively on the angle between the molecular axis and the laser polarization.\cite{endo2022post,dimitrovski2011ionization,holmegaard2010photoelectron} 

For the S2 channel, the total kinetic-energy distribution spans $\approx$2.8–7.1 eV, with a peak centered at approximately 4.5 eV. A notable feature of this distribution is its close correspondence to the distribution of the C2 channel of the CO$^+$ fragment as illustrated in Figure~\ref{Figure 5} (see also Section~\ref{sec:co+}). The near-identical positions and shapes of these features suggest that both arise from the same dissociation pathway, namely the Coulomb explosion of OCS$^{2+}$ yielding $\mathrm{CO}^+ + \mathrm{S}$$^+$. Kinetic-energy distributions often serve as fingerprints of the underlying dissociative electronic states, and comparison with prior photoionization imaging and coincidence studies provides an effective basis for assigning the S2 channel. 

Previous experimental and theoretical work has shown that OCS$^{2+}$, once formed by various ionization schemes, fragments predominantly to $\mathrm{CO}^+ + \mathrm{S}$$^+$, with only minor branching to CS$^+$ + O$^+$ or SO$^+$ + C$^+$.\cite{zhao2019strong, brites2008ocs2+}. Investigations by Jarraya \textit{et al.}\cite{jarraya2023doubly} and Endo \textit{et al.}\cite{endo2022post} have revealed that the formation of the $\mathrm{CO}^+ + \mathrm{S}$$^+$ ion pair proceeds via two distinct dissociation pathways: a dominant high-energy channel with an appearance energy of (34.1 ± 0.5)~eV, associated with KERs of approximately 5.5 eV, and a secondary low-energy channel with an appearance energy of (31.7 ± 0.4)~eV, characterized by KERs of approximately 4 eV. Photoion–photoion coincidence (PIPICO) measurements in the 37–100 eV photon-energy range further showed that, for the $\mathrm{CO}^+ + \mathrm{S}$$^+$ channel at $h\nu$ = 37 eV, the KER spans 3.0 – 6.3~eV with an average of 4.5 eV.\cite{masuoka1992kinetic} Likewise, velocity-map imaging studies of dissociative electron-impact ionization (50–100 eV) reported a total KER of $\sim$ (6.0 ± 0.4)~eV for the high-energy $\mathrm{CO}^+ + \mathrm{S}$$^+$ channel.\cite{lomas2024multimass}
Threefold and fourfold electron–ion coincidence spectroscopy at 40.81 eV,\cite{jarraya2023doubly} supported by quantum-chemical calculations, revealed two distinct dissociation pathways of OCS$^{2+}$ leading to $\mathrm{CO}^+ + \mathrm{S}$$^+$:
(1) a lower-energy channel corresponding to direct fragmentation of $\mathrm{OCS}^{2+}(X^3\Sigma^-)$, producing $\sim$ 4 eV of total kinetic energy and dissociating to the lowest asymptote $\mathrm{CO}^+(X\,{}^{2}\Sigma^{+}) + \mathrm{S}^+({}^{4}S)$; 
and (2) a higher-energy channel near 5.2 eV which is accessed following $\mathrm{OCS}^{2+}$ → $\mathrm{COS}^{2+}$ isomerization proceeding via $\mathrm{OCS}^{2+}(b\,{}^{1}\Sigma^{+})$ → $\mathrm{COS}^{2+}(X\,{}^{3}\Sigma^{-})$
 → $\mathrm{CO}^+(X\,{}^{2}\Sigma^{+}) + \mathrm{S}^+({}^{4}S)$.
 
The S2 feature spanning total kinetic energies from 2.8 to 7.1 eV with a peak at approximately 4.5 eV measured in the present work is consistent with these earlier findings. The dominant peak near 4 eV suggests that direct fragmentation of $\mathrm{OCS}^{2+}(X^3\Sigma^-)$ provides the principal contribution to S2, while the weaker high-energy shoulder near 5.2 eV may reflect a minor population of the isomerization-mediated fragmentation channel.

The appearance of channel S3 only at high laser power suggests that it does not originate from dissociation of OCS$^+$. In addition, the KER distributions of S3 differ markedly from those of the CO$^+$ producing channels, indicating that S3 is unlikely to arise from the Coulomb explosion pathway OCS$^{2+}$ → $\mathrm{CO}^+ + \mathrm{S}$$^+$. 

Using synchrotron radiation in combination with photoion–photoion coincidence (PIPICO) techniques \cite{masuoka1992kinetic}, KERs of fragment ions produced in the dissociative double photoionization of OCS have been extensively investigated. Three dominant dissociation channels of the OCS$^{2+}$ ion were found: $\mathrm{CO}^{+} + \mathrm{S}^{+}$, $\mathrm{C} + \mathrm{S}^{+} + \mathrm{O}^{+}$, and $\mathrm{C}^{+} + \mathrm{S}^{+} + \mathrm{O}$. 

Photoelectron–photoion–photoion coincidence (PEPIPICO) measurements by Eland\cite{eland1987dynamics} revealed that dissociation of OCS$^{2+}$ into two charged and one neutral fragment can proceed via distinct mechanisms. In particular, the $\mathrm{C}^{+} + \mathrm{S}^{+} + \mathrm{O}$ channel was shown to arise from a sequential dissociation process, $\mathrm{OCS}^{2+} \rightarrow \mathrm{CO}^{+} + \mathrm{S}^{+} \rightarrow \mathrm{C}^{+} + \mathrm{S}^{+} + \mathrm{O}$, whereas the formation of $\mathrm{C} + \mathrm{S}^{+} + \mathrm{O}^{+}$ fragments was attributed to an essentially instantaneous three-body breakup. Subsequent electron-impact ionization studies by Wang \textit{et al}.\cite{wang2003dissociation} and Shen \textit{et al}.\cite{shen2016fragmentation} further supported a sequential dissociation mechanism for the $\mathrm{C}^{+} + \mathrm{S}^{+} + \mathrm{O}$ channel. 

Complementary velocity-map imaging studies of OCS$^{2+}$ produced by electron-impact ionization by Lomas \textit{et al.}\cite{lomas2024multimass} revealed signatures of an initial charge-separation step followed by delayed fragmentation. The corresponding KER distributions were found to exhibit a pronounced high-energy component ($\sim$10\,eV) associated with the $\mathrm{CO}^{+} + \mathrm{S}^{+}$ charge separation, together with a near-zero KER for the subsequent loss of neutral oxygen from the CO$^{+}$ intermediate.\cite{lomas2024multimass} For the $\mathrm{C} + \mathrm{S}^{+} + \mathrm{O}^{+}$ channel, the results were consistent with a concerted three-body dissociation mechanism.

In summary, these previous results suggest that the lowest energy pathways leading to the formation of channel S3 of the S$^{+}$ product are predominantly associated with three-body dissociation processes, encompassing both a sequential charge-separation pathway $\mathrm{C}^{+} + \mathrm{S}^{+} + \mathrm{O}$ proceeding via the $\mathrm{CO}^{+} + \mathrm{S}^{+}$ intermediate and a concerted fragmentation channel yielding $\mathrm{C} + \mathrm{S}^{+} + \mathrm{O}^{+}$ products. 
Because VMI images of the C/C$^+$ and O/O$^+$ coproducts were not recorded in the present study, further mechanistic distinctions between these possibilities cannot be established at present.

\subsection{CO$^+$ channel}
\label{sec:co+}

In the image corresponding to the CO$^+$ mass channel shown in Figure~\ref{fgr:Figure 6}, three distinct dissociation channels, denoted C1, C2, and C3, are identified. At low laser intensity, C1 is the dominant product channel, while C2 appears as a minor contribution. With increasing laser intensity, a third channel, C3, becomes pronounced. All three channels exhibit strongly anisotropic angular distributions (Figure S4 in the ESI), indicating that they arise from prompt, direct dissociation processes occurring on ultrafast timescales.

A key observation is that, as discussed in the previous section, the CO$^+$-producing channel C2 coincides with the S2 channel observed in the S$^+$ image (Figure~\ref{Figure 5}), demonstrating that these two fragment ions originate from the same dissociation pathway, i.e., the Coulomb explosion of doubly charged OCS$^{2+}$, yielding correlated $\mathrm{CO}^+ + \mathrm{S}$$^+$ fragment pairs.

The dominance of C1 at low laser intensity suggests that this channel primarily results from dissociation of singly charged OCS$^+$ via the pathway OCS$^+$ → S + CO$^+$. The corresponding total kinetic-energy distribution spans from 0 to 2.1 eV and exhibits a pronounced peak at approximately 0.47 eV. Figure~\ref{fgr:Figure 3} suggests that the lowest accessible electronically excited state that is capable of dissociating into S + CO$^+(X^{2}\Sigma^{+})$ is the $C^{2}\Sigma^{+}$ state.\cite{hubin1996dissociation} The electron-impact ionization experiments by Lomas \textit{et al.} showed that dissociation of singly charged parent ions contributes to CO$^+$ velocity-map images with the kinetic energy distribution peaking below 0.5 eV and extending to approximately 2 eV.\cite{lomas2024multimass} This behavior was discussed to be consistent with dissociation proceeding predominantly via population of the $C^{2}\Sigma^{+}$ state of OCS$^+$, followed by intersystem crossing or internal conversion to repulsive $^{2,4}\Sigma^{-}$ or $^{2,4}\Pi$ states connecting to the $\mathrm{S}({}^{3}P) + \mathrm{CO}^+(X\,{}^{2}\Sigma^{+})$ asymptote. 

The emergence of channel C3 exclusively at high laser intensities suggests that it does not originate from dissociation of singly charged OCS$^+$. Furthermore, the KER distributions associated with C3 differ substantially from those of the S$^+$ producing channels, suggesting that C3 is unlikely to result from the Coulomb explosion pathway OCS$^{2+}$ → $\mathrm{CO}^+ + \mathrm{S}$$^+$. Instead, the lowest-energy processes capable of producing CO$^+$ in this channel are most consistent with dissociation of triply charged OCS$^{3+}$ via the pathway OCS$^{3+}$ → $\mathrm{CO}^+ + \mathrm{S}^{2+}$.\cite{masuoka1991dissociative} Because velocity-map imaging data for the $\mathrm{S}^{2+}$ coproduct were not acquired, a more definitive assignment of the underlying dissociation mechanism cannot be made at present.

\section{Conclusion}

In this work, we have systematically characterized the fragmentation dynamics of OCS under strong-field-ionization conditions, supported by ab-initio PES, with particular emphasis on the distinct roles of singly and doubly charged species. 
For OCS$^+$, the dominant dissociation pathways were identified as two-body channels leading to S$^+$ + CO (channel S1) and S + CO$^+$ (channel C1), with KER distributions that can be directly rationalized in terms of population of the low-lying X$^2\Pi$, A$^2\Pi$, B$^2\Sigma^+$, $C^{2}\Sigma^{+}$ electronic states and their associated dissociation limits. These channels exhibit relatively low KER and reflect intramolecular bond cleavage governed by the underlying cationic PESs.

In contrast, fragmentation of OCS$^{2+}$ is dominated by charge separation, giving rise to the CO$^+$ + S$^+$ channel (channels S2 and C2) with significantly higher KER. The observed KER distributions are consistent with dissociation from both the lowest bound states and higher-lying repulsive regions of the OCS$^{2+}$ PESs, including contributions from direct dissociation and isomerization-mediated pathways. Furthermore, the emergence of higher-energy features (channels S3 and C3), assigned to three-body fragmentation and may involving triply charged OCS$^{3+}$, highlights the increasing importance of multichannel dynamics at elevated laser intensities.

The momentum-resolved measurements further reveal clear differences between monomer and dimer fragmentation. While monomer dynamics are governed by intramolecular charge localization, the dimer signals exhibit distinct signatures, including low-energy features (channel R1), and high-energy Coulomb explosion contributions (channel R2 and R3). These observations are attributed to fragmentation of multiply charged OCS dimers via competing charge-partition pathways, such as (OCS)$_2^{2+}$, (OCS)$_2^{3+}$, and (OCS)$_2^{4+}$, leading to charge-separated dissociation.

In summary, the present results illustrate how dynamic signatures of strong-field fragmentation evolve from intramolecular dissociation in isolated molecules to intermolecular charge separation in weakly bound clusters, thus yielding a unified picture of charge-driven dissociation dynamics beyond the single-molecule limit.

\section*{Conflicts of interest}
There are no conflicts to declare.

\section*{Data availability}
The primary data supporting the findings of this article are available at Zenodo at  https://doi.org/10.5281/zenodo.20736015.

\section*{Acknowledgements}
We thank Dr A. Johnson, G. Holderried, D. Baumgartner, G. Martin, and Ph. Knöpfel for technical assistance. This research was supported by the Swiss National Science Foundation (grant nr. TMAG-2\_209193) and the University of Basel.

\clearpage

\renewcommand{\thetable}{S\arabic{table}}
\renewcommand{\thefigure}{S\arabic{figure}}
\renewcommand{\thesection}{S\arabic{section}}
\renewcommand{\d}{\text{d}}
\setcounter{figure}{0}  
\setcounter{section}{0}  
\setcounter{table}{0}

\noindent
{\centering
\Large\textbf{Supplementary Information} \\[1em]
\par}

\begin{figure}[H]
\centering
\includegraphics[scale=0.11]{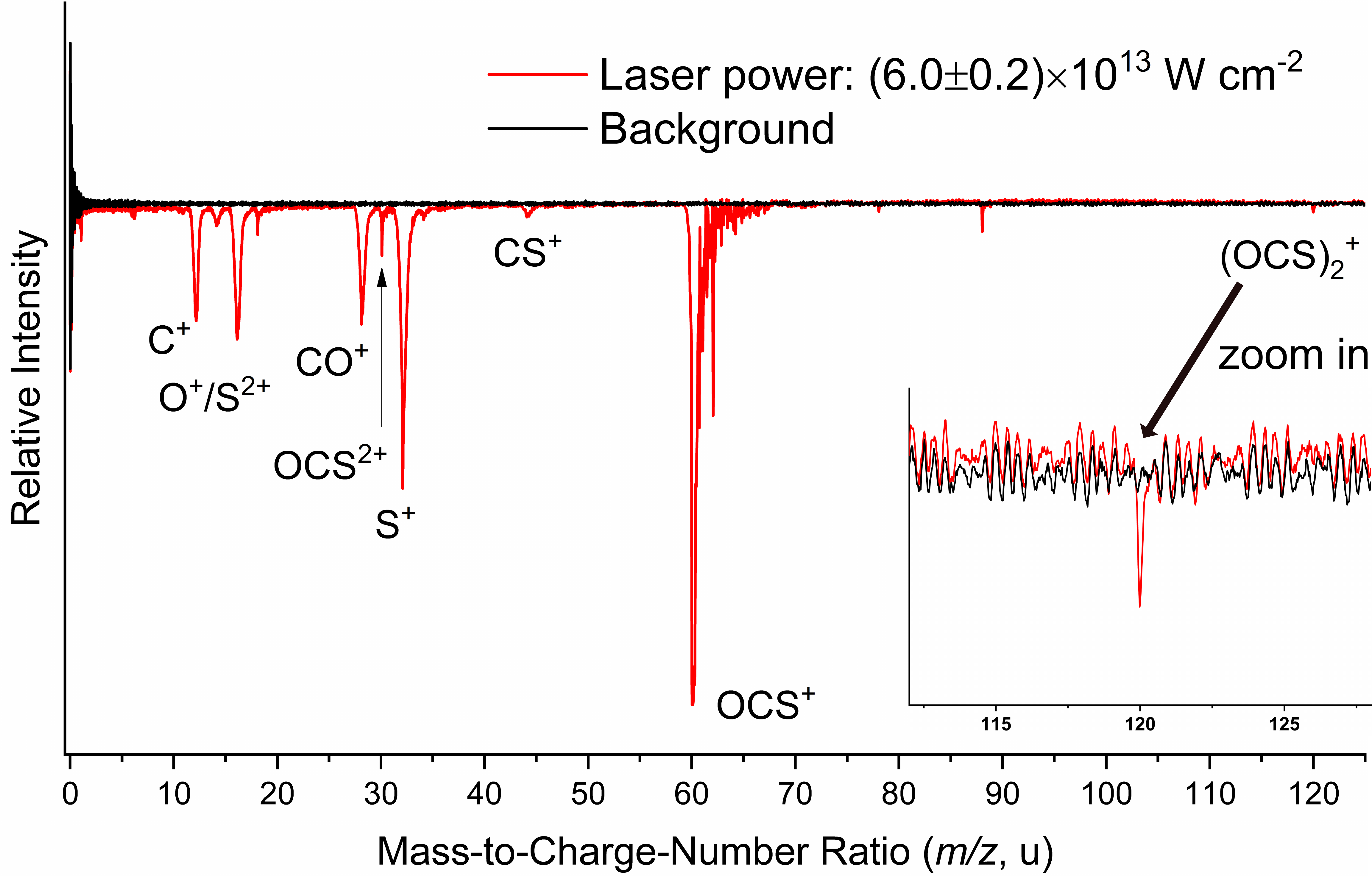} 
\caption{Time-of-flight mass spectra of ionic products generated from 1000 ppm carbonyl sulfide (OCS) seeded in helium (He) by ionization using femtosecond laser pulses (775 nm, 150 fs) at a peak intensity of $(6.0 \pm 0.2)\times 10^{13}\ \mathrm{W,cm^{-2}}$ (red trace). The black trace represents the background signal. The inset highlights the ion signal at m/z = 120~u.}
\label{sifig:Figure S1}
\end{figure}

\begin{figure}[H]
\centering
\includegraphics[scale=0.45]{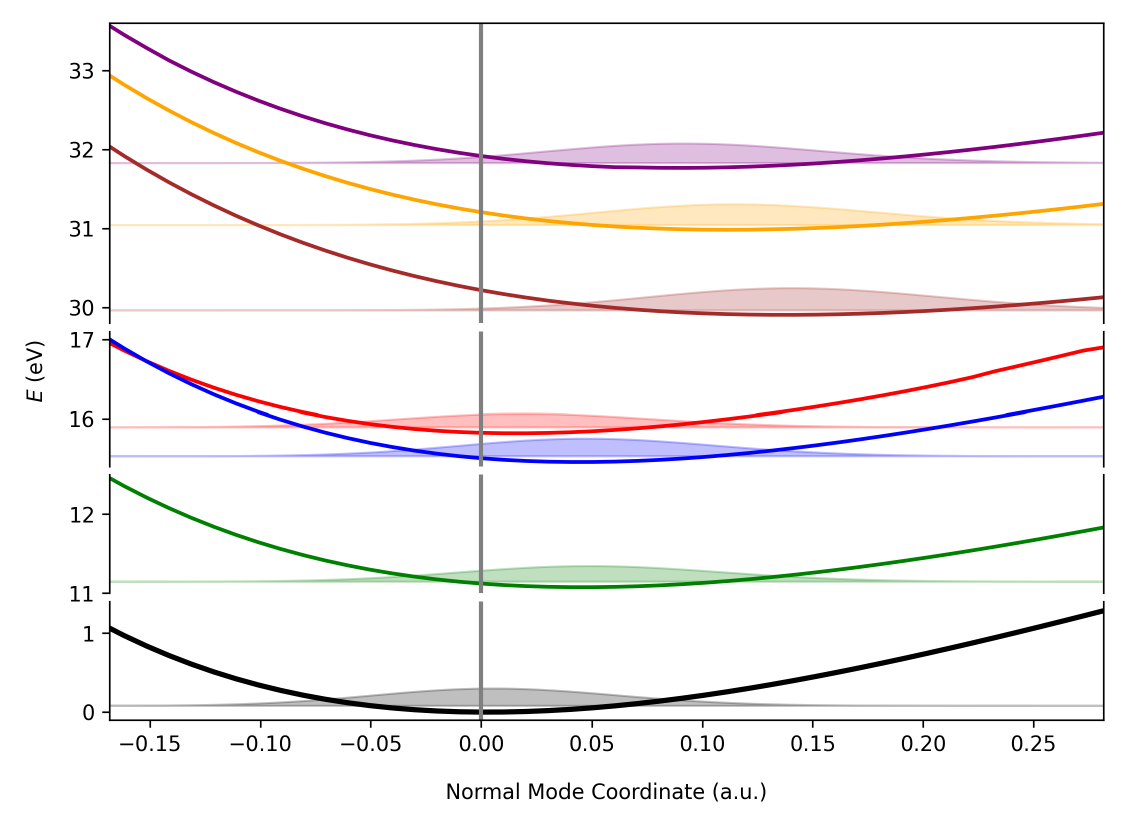} 
\caption{Comparison of potential energy curves of OCS and its cations: neutral ground state (black: X$^1\Sigma^+$), three lowest-lying electronic states of OCS$^+$ (green: X$^2\Pi$, blue: A$^2\Pi$, red: B$^2\Sigma^+$), three lowest-lying electronic states of OCS$^{2+}$ (brown: X$^3\Sigma^-$, orange: a$^1\Delta$, purple: b$^1\Sigma^+$). The X-axis represents the C-S stretching normal coordinate. The vertical gray line marks the equilibrium structure of neutral OCS. In addition, the ground-state vibrational wavefunctions for each PES are also depicted as shaded areas, colored to match their corresponding PES. The equilibrium geometry of OCS was optimized at the MRCI/aVTZ level with Davidson-relaxed reference correction, while the normal mode vectors were obtained using CCSD(T)/aVTZ. 
}
\label{sifig:Figure S2}
\end{figure}

\begin{table}
\small
  \caption{\ Franck–Condon factors for transitions from the ground vibrational state of neutral OCS in its X$^1\Sigma^+$ ground electronic state to vibrational states $v$ in selected electronic states of OCS$^+$ and OCS$^{2+}$.}
  \label{tbl:FC}
  \begin{tabular}{lllllllll}
    \hline
   State & $v$=0 & $v$=1 & $v$=2 & $v$=3 & $v$=4 & $v$=5 & $v$=6 & $v$=7 \\
    \hline
    OCS$^+$ (X$^2\Pi$) & 0.75 & 0.22 & 0.03 \\
    OCS$^+$ (A$^2\Pi$) & 0.75 & 0.20 & 0.04 & 0.01 \\
    OCS$^+$ (B$^2\Sigma^+$) & 0.97 & 0.03 \\
    OCS$^{2+}$ (X$^3\Sigma^-$) & 0.08 & 0.19 & 0.24 & 0.21 & 0.15 & 0.08 & 0.04 & 0.01 \\
    OCS$^{2+}$ (a$^1\Delta$) & 0.18 & 0.29 & 0.25 & 0.16 & 0.08 & 0.03 & 0.01 \\
    OCS$^{2+}$ (b$^1\Sigma^+$) & 0.33 & 0.35 & 0.20 & 0.08 & 0.03 & 0.01 \\
    \hline
  \end{tabular}
\end{table}

\begin{figure}[H]
\centering
  \includegraphics[height=3.5cm]{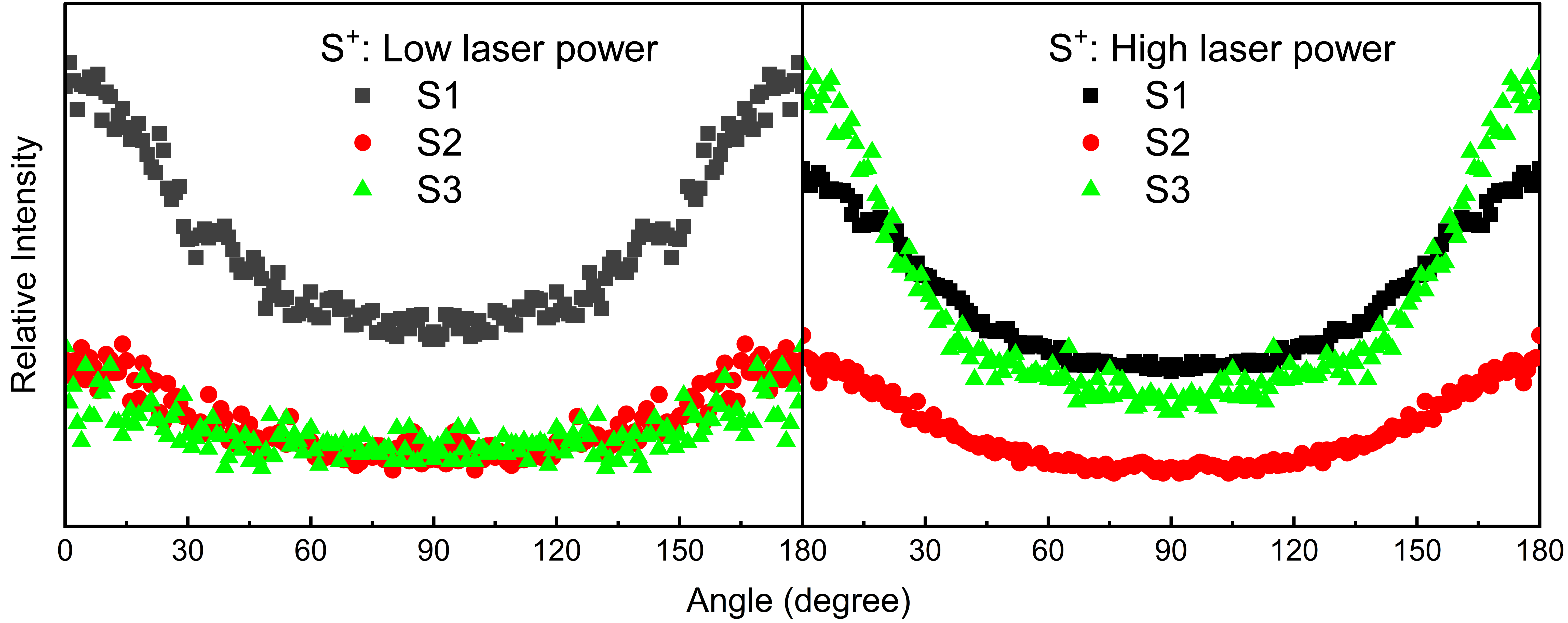}
  \caption{Angular distribution of S1, S2, S3 channels of product S$^+$ at low (left) and high (right) laser power.}
  \label{fgr:Figure S3}
\end{figure}

\begin{figure}[H]
\centering
  \includegraphics[height=3.5cm]{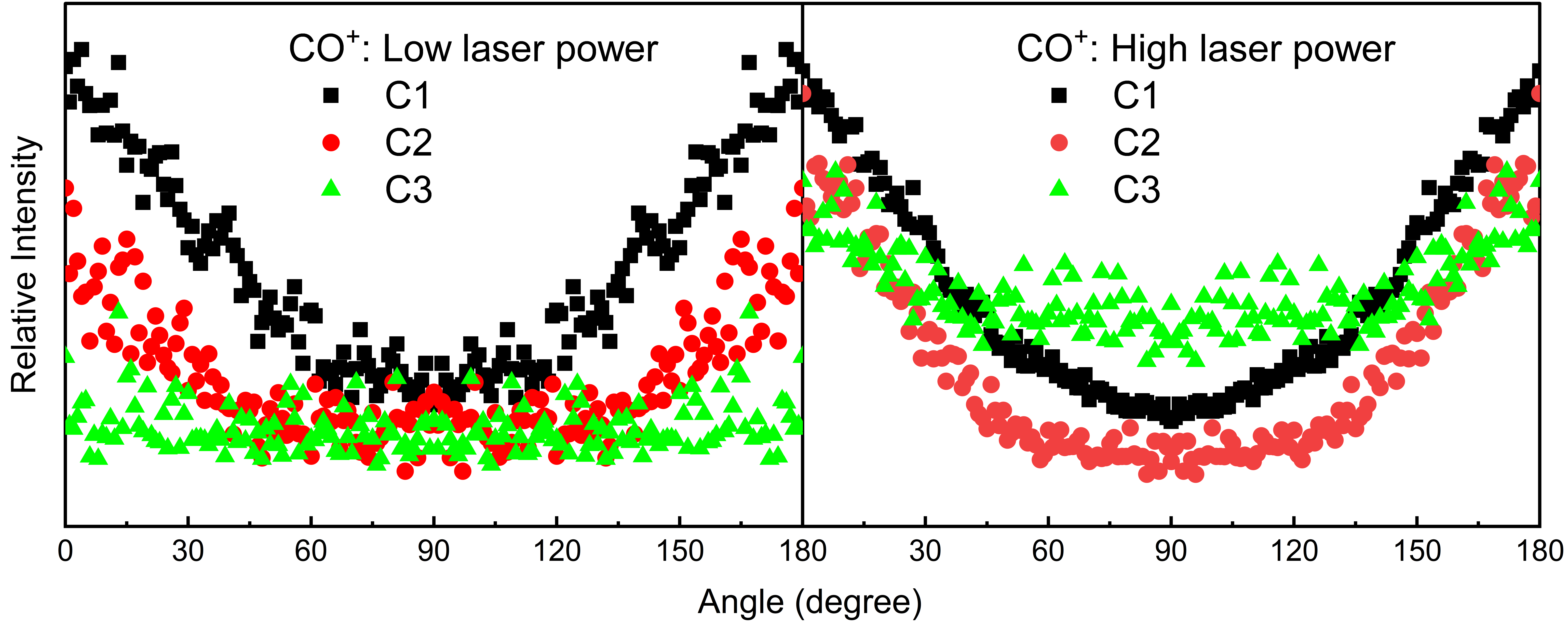}
  \caption{Angular distribution of product channels C1, C2, C3 of CO$^+$ at low (left) and high laser power (right).}
  \label{fgr:Figure S4}
\end{figure}

\balance

\clearpage

\bibliography{rsc} 
\bibliographystyle{rsc}

\end{document}